\documentclass[preprint]{aastex}
\slugcomment{{\sc Accepted for publication in the Astronomical Journal}}
\shortauthors{Krist et al.}
\shorttitle{HD 207129}

\def\asec{$^{\prime\prime}$~}

\def\amin{$^{\prime}$~}
\def\Msun{M$_{\odot}$~}
\def\Lsun{L$_{\odot}$~}

\def\deg{$^{\circ}$~}

\def\gapp{\lower 3pt\hbox{${\buildrel > \over \sim}$}\ }

\begin{document}

\title{{\it HST} and {\it Spitzer} Observations of the HD 207129 Debris Ring}

\author{John E. Krist\altaffilmark{1}, Karl R. Stapelfeldt\altaffilmark{1},
Geoffrey Bryden\altaffilmark{1,2}, George H. Rieke\altaffilmark{3},  
K. Y. L. Su\altaffilmark{3}, Christine C. Chen\altaffilmark{4},
Charles A. Beichman\altaffilmark{5}, Dean C. Hines\altaffilmark{6},
Luisa M. Rebull\altaffilmark{7}, Angelle Tanner\altaffilmark{8},
David E. Trilling\altaffilmark{9}, Mark Clampin\altaffilmark{10},
Andr\'as G\'asp\'ar\altaffilmark{3}
}

\altaffiltext{1}{ Jet Propulsion Laboratory, California Institute of 
Technology, 4800 Oak Grove Drive, Pasadena CA 91109 }
\altaffiltext{2}{ NASA Exoplanet Science Institute, 
California Institute of Technology, 770 S Wilson Ave, Pasadena, CA 91125 }
\altaffiltext{3}{ Steward Observatory, University of Arizona, 
  933 N Cherry Ave, Tucson, AZ 85721 }
\altaffiltext{4}{ Space Telescope Science Institute, 3700 San Martin Drive, Baltimore MD 21218 }
\altaffiltext{5}{ NASA Exoplanet Science Institute, California Institute of Technology, 770 S. Wilson Ave., Pasadena CA 91125 } 
\altaffiltext{6}{ Space Science Institute, 4750 Walnut St. Suite 205, Boulder CO 80301 }
\altaffiltext{7}{ Spitzer Science Center, Mail Stop 220-6, California Institute of Technology, Pasadena CA 91125 }
\altaffiltext{8}{ Georgia State University, Department of Physics and Astronomy, One Park Place, Atlanta GA 30316  }
\altaffiltext{9}{ Department of Physics and Astronomy, Northern Arizona University, Box 6010, Flagstaff AZ 86011 }
\altaffiltext{10}{ NASA Goddard Space Flight Center, Greenbelt, MD 20771}

\begin{abstract}

A debris ring around the star HD 207129 (G0V; $d$ = 16.0 pc) has been imaged in
scattered visible light with the ACS coronagraph on the {\it Hubble Space
Telescope} and in thermal emission using MIPS on the {\it Spitzer Space
Telescope} at $\lambda$ = 70 $\mu$m (resolved) and 160 $\mu$m (unresolved).
{\it Spitzer} IRS ($\lambda$ = 7--35 $\mu$m) and MIPS ($\lambda$ = 55--90
$\mu$m) spectrographs measured disk emission at $\lambda >$ 28 $\mu$m.  In the
$HST$ image the disk appears as a $\sim$30 AU wide ring with a mean radius of
$\sim$163 AU and is inclined by 60\deg from pole-on.  At 70 $\mu$m it appears
partially resolved and is elongated in the same direction and with nearly the
same size as seen with $HST$ in scattered light.  At 0.6 $\mu$m the ring shows
no significant brightness asymmetry, implying little or no forward scattering
by its constituent dust.  With a mean surface brightness of $V=$
23.7 mag arcsec$^{-2}$, it is the faintest disk imaged to date in scattered
light.  

We model the ring's infrared spectral energy distribution using a dust 
population fixed at the location where $HST$ detects the scattered light.
The observed SED is well-fit by this model, with no requirement for 
additional unseen debris zones.  The firm constraint on the dust radial
distance breaks the usual grain size-distance degeneracy that exists in
modeling of spatially unresolved disks, and allows us to infer a minimum 
grain size of $\sim 2.8 \mu$m and a dust size distribution power law spectral 
index of -3.9.  An albedo of $\sim$5\% is inferred from the
integrated brightness of the ring in scattered light.  The low albedo
and isotropic scattering properties are inconsistent with Mie theory for 
astronomical silicates with the inferred grain size and show the need
for further modeling using more complex grain shapes or compositions.

Brightness limits are also presented for six other main sequence stars with
strong {\it Spitzer} excess around which $HST$ detects no circumstellar 
nebulosity (HD 10472, HD 21997, HD 38206, HD 82943, HD 113556, and HD 138965).

\end{abstract}

\vfil\eject

\section{Introduction}

Circumstellar debris disks are created by the collisions or disruptions of
solid bodies (asteroids, comets, planets) that generate clouds of small dust
grains.  Radiation pressure, Poynting-Robertson drag, and stellar winds can
remove these particles on timescales much less than the age of the star.  Thus,
seeing a debris disk means that either a major collision of large objects has
recently occurred in the system, or that the dust is continually replenished
through collisions within a reservoir of smaller bodies.  Either way, these
disks signify the presence of some kind of planetary system.  They also offer
the chance to view a system of planetesimals for signs of unseen planets or
very-low-mass companions by way of central clearings, gaps, localized clumps,
or spiral arms caused by planetary perturbations on the disk structure.

An increasing number of debris disks are being resolved, especially around
solar-type stars.  Infrared measurements from the {\it IRAS}, {\it ISO}, and
{\it Spitzer} space telescopes have been used to identify disk candidates based
on far-infrared flux densities well in excess of those expected for stellar
photospheres.  A few have been resolved in long-wavelength emission with {\it
Spitzer} and submillimeter radio telescopes.  Such disks, like those around
$\beta$ Pictoris (Heap et al. 2000) or Fomalhaut (Kalas et al. 2005), have low
optical depths that make them difficult to image in scattered light relative to
the glare created by the stars and telescope.  High contrast imaging techniques
are required, including a coronagraph to suppress the diffraction pattern of
the star and point spread function (PSF) subtraction to reduce the residual
instrumentally-scattered light.  While ground-based observations use these
methods, the lack of wavefront stability due to atmospheric turbulence, even
after correction by adaptive optics, precludes them from detecting all but the
brightest debris disks. This is largely due to image instabilities that lead to
significant PSF subtraction residuals.  The {\it Hubble Space Telescope} ({\it
HST}) avoids this problem entirely and provides a highly stable wavefront,
making it the premier tool for debris disk imaging at visible wavelengths.  
The coronagraph on the Advanced Camera for Surveys (ACS) provided the highest
contrast imaging capability on {\it HST} (Krist 2004) until its failure in
early 2007.

By showing the location of orbiting planetesimals and, in some cases, by
identifying disk asymmetries caused by individual planets, debris disk images
probe the architecture of planetary systems around other stars.  For the most
direct comparison with the Solar System, debris disks around mature solar-type
stars are of particular interest.  We consider here the nearby Sun-like star HD
207129 (HR 8323, HIP 107649, GJ 838, IRAS 21450-4732; $V$=5.58), a G0V star at
a Hipparcos-measured distance of 16.0 pc (van Leeuwen 2007).  Walker \&
Wolstencroft (1988) noted that HD 207129 had a small {\it IRAS} 60 $\mu$m
excess while Habing et al.  (1996) measured 60 and 90 $\mu$m {\it ISO}
excesses, indications of circumstellar material, most likely a disk.  Jourdain
de Muizon et al. (1999) obtained additional {\it ISO} observations over
$\lambda$ = 2.5 -- 180 $\mu$m and verified that there was no significant excess
below 25 $\mu$m.  This indicates the lack of circumstellar material near the
star and thus a central clearing in the disk.  They suggested that a substellar
companion interior to the clearing may be required to maintain the hole as
Poynting-Robertson drag would cause dust to migrate towards the star and fill
it on short timescales.  Sheret, Dent, \& Wyatt (2004) were unable to detect
the disk at $\lambda$ = 450 and 850 $\mu$m using SCUBA.  Using the
submillimeter non-detections as upper limits and combining the {\it IRAS} and
{\it ISO} results, they derived a disk radius of $260\pm50$ AU.  However,
Zuckerman \& Song (2004) derived a 35 AU orbital radius by assuming that large
dust grains emit as a blackbody at T$\sim$50 K.  This highlights a problem with
determining disk sizes from unresolved photometric measurements: degeneracies
exist among the grain size, emissivity, temperature, and distance from the
star, and thus the size scale of the disk is uncertain. For example, see the
analysis of the HD 12039 disk by Hines et al. (2006).  By directly measuring
the location of the dust, resolved images of the disk can break these
degeneracies and allow reliable grain properties to be derived. 

In an effort to increase the number of resolved debris disks, we utilized the
ACS coronagraph to image disk candidate stars selected based on their infrared
excesses measured with the {\it IRAS}, {\it ISO}, and {\it Spitzer} space
telescopes.  Most disk candidates with higher optical depths (implied by
$L_{\rm dust}/L_{\star} \gapp 5\times10^{-4}$) have previously been imaged by 
{\it HST} and ground-based telescopes.  To date, the lowest optical depth disk 
seen in scattered light is Fomalhaut's (Kalas et al. 2005), which has
$L_{\rm dust}/L_{\star} = 8\times10^{-5}$.  We chose eight targets
previously unobserved by {\it HST} having $L_{\rm dust}/L_{\star}$ = 1--7
$\times10^{-4}$, including HD 207129.  We report here the detection of a disk
around that star and, as an appendix, non-detections around six of the other
targets (the detection of the disk around HD 10647 is discussed in Stapelfeldt
et al. in preparation).

\section{OBSERVATIONS AND DATA PROCESSING}

\subsection{{\it HST} ACS Observations}

HD 207129 was observed on 3 May 2006 with the coronagraph in the ACS High
Resolution Camera (HRC; $\sim$0\farcs 025 pixel$^{-1}$) in {\it HST} program
10539.  Two observation sequences were executed, each consisting of a 0.1s
exposure in the narrowband filter F502N for automated acquisition and centering
of the star behind the coronagraphic occulting spot, and a 100s exposure and
four 520s exposures in F606W with the star centered behind the 1\farcs
8-diameter occulter.  At the beginning of the first sequence, two 0.3s direct
(non-coronagraphic) exposures in filter F606W (ACS wide-$V$ band) were taken to
provide unsaturated images for accurate stellar photometry.  Each sequence
required one orbit, and the two sequences were executed in consecutive orbits.
Between them, the telescope was rolled 20.0\deg about the line of sight to
provide additional discrimination between real objects and instrumental
artifacts (instrumental artifacts appear stationary on the detector while the
sky image rotates).

While the coronagraph suppresses the diffraction pattern created by the
telescope, it does not reduce the scattering by optical surface errors that
create a halo of light around the star.  This halo is typically subtracted
using an image of another isolated star observed in the same manner.  This is
practical because the {\it HST} PSF is generally stable over time (compared to
ground-based telescopes).  The G3V star HD 211415 ($V$=5.33) was observed in
the orbit following the two HD 207129 orbits to provide a reference PSF. It was
selected because it has a similar color to HD 207129 and is nearby in the sky.
Unfortunately our images showed the presence of a companion star located
2\farcs 2 away at PA=43.2\deg and 4.5 mag fainter\footnote{Jasinta, Raharto,
and Soegiartini (1995) also report this object with separation of 4.96\asec at
PA 37.8\deg, and identical brightness relative to the primary star.  Given 
HD 211415's large proper motion of 771 mas/yr at PA 124.8\deg (Perryman et al.
1997), this must be a bound companion star in a nearly edge-on orbit with
semi-major axis $\sim$ 100 AU.}.  In the coronagraphic images, the companion is
highly saturated, rendering these images unsuitable for use for PSF
subtraction.  As will be described later, an alternative subtraction method was
used instead.

The ACS images were calibrated by the {\it HST} pipeline.  Multiple long
exposures were combined and cosmic rays rejected.  Image count rates were
converted to standard $V$ magnitudes using the zero point obtained from the
SYNPHOT synthetic photometry program based on a stellar spectrum of similar
type to HD 207129.

\subsection{{\it HST} ACS PSF Subtraction}

As previously noted, the designated PSF subtraction reference star, HD 211415,
was a binary.  To replace it, ten similarly-observed alternative reference
stars were collected from our program and the $HST$ Archive.  Each was
subtracted from the combined long exposure HD 207129 images.  This was done by
iteratively shifting the reference image via interpolation, adjusting its
normalization, and subtracting it from the science image until the residual
instrumental halo was minimized by eye.  None of these subtractions indicated
the presence of any circumstellar material above the level of the residuals,
and they are therefore of no further use.  Using these reference stars was not
optimal because the telescope focus and the coronagraphic occulter position are
known to vary over time, and this selection of reference stars was not matched
in color with HD 207129 as HD 211415 had been.  These differences between the
PSF reference stars and the science target can cause significant subtraction
residuals.  

The only viable alternative processing technique was to use the image of HD
207129 taken at one orientation to subtract the starlight from the image taken
at the other and vice-versa, a process called roll subtraction.  Because the
images were taken close in time, this reduces the effects of time-dependent
changes in the telescope, and there is no color mismatch.  However, any
extended disk could subtract part of itself in the other orientation, thus this
method is best suited to nebulosity that is confined along a preferred axis
through the star, such as an edge-on disk (a symmetric face-on disk would
completely subtract itself, for example).

A roll-subtracted image contains a positive image of a disk and a negative one
rotated by an angle equal to the difference between the telescope orientations.
Ideally, these two would be combined to form a single, positive image, but that
is difficult to do after subtraction.  Instead of directly subtracting images
from the two orientations and then trying to combine the results, an iterative
technique can be used that solves for those portions of the two unsubtracted
images that appear static on the detector ($i.e.$\ the PSF) and those that 
appear to rotate as the telescope rolls ($i.e.$ the sky, including any disk).  
Such a method was used to subtract the {\it HST} STIS images of $\beta$ 
Pictoris (Heap et al. 2000), whose edge-on disk is well-suited for this.  

The procedure begins with an initial estimate of the static (PSF) component,
which is the average of the unsubtracted (raw) images (it is assumed
that the PSF is largely stable between the two observations).  At this point,
any rotating (sky) features from each orientation in this static image are now
half of their original intensities compared to the total static component.
This estimate of the static component is then subtracted from each raw image, 
creating the initial sky estimate at each orientation.  One of the sky frames 
is then
rotated to the orientation of the other, and the two are then averaged.  This
new sky frame is then subtracted from each raw frame (after rotation), and the
result is averaged, creating a new static estimate.  In each subsequent 
iteration, the residual sky features in the estimated static (PSF) component
are reduced in intensity by a factor of two.  This process is repeated
until the sky frame does not visibly change.

The above method was applied to the long-exposure HD 207129 data.  Before doing
so, the images from the two roll angles were shifted via interpolation to
a common center, to remove small residual pointing offsets between the
two.  The subtraction procedure was then run for eight iterations.  The result 
revealed a very faint, inclined ring centered on the star.  To improve the 
signal, the original images were downsampled to 0\farcs 1 resolution using 
4$\times$4 rebinning and the subtraction process repeated with improved 
results (Figure 1).

\begin{figure}
\epsscale{0.5}
\plotone{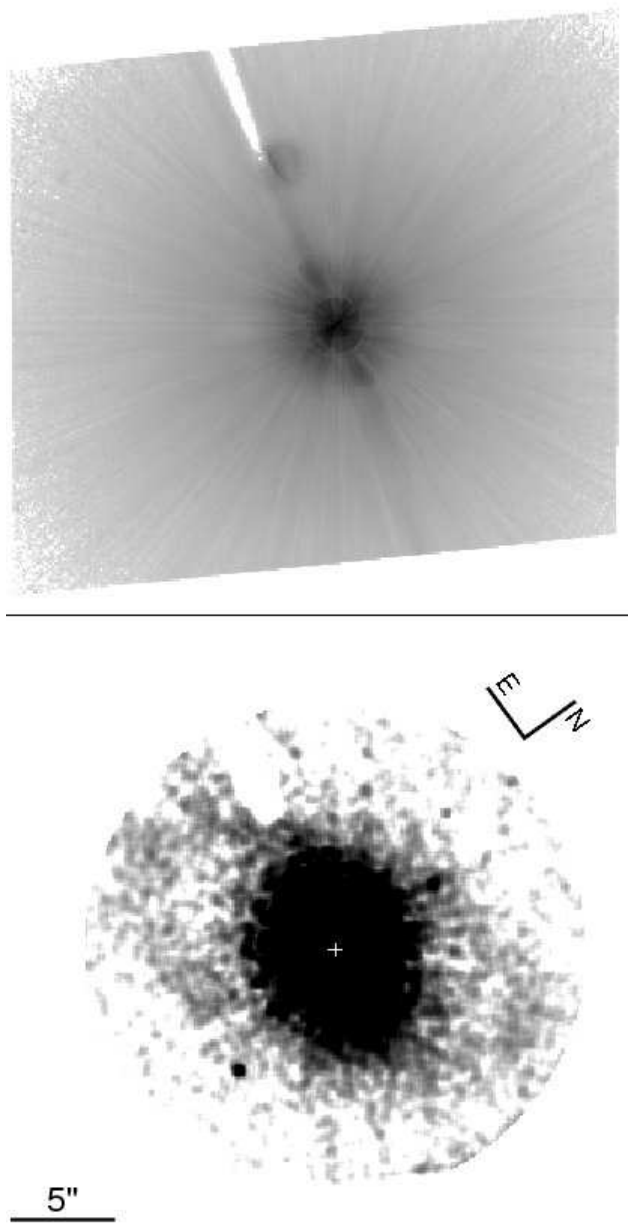}
\caption{ACS F606W coronagraphic observations of HD 207129. Each frame is
30\farcs 2 on a side. (TOP) Coronagraphic image from the first orientation
prior to PSF subtraction.  Shadows of the occulting finger and larger occulting
spot are seen towards the upper left corner.  The diagonal streak stretching
from the upper left to lower right is instrumentally-scattered light.  This
image is displayed with logarithmic intensity scaling. (BOTTOM) Image of the HD
207129 ring after applying the iterative roll subtraction method described in
the text to the exposures taken at each orientation.  The images were rebinned
to 0\farcs 1 sampling prior to processing, and the result was additionally
smoothed using a median filter.  This image is displayed with a quarter-root
intensity scaling with a much lower maximum value compared to the top image. }
\epsscale{1.0}
\end{figure}

\begin{figure} 
\epsscale{0.7} 
\plotone{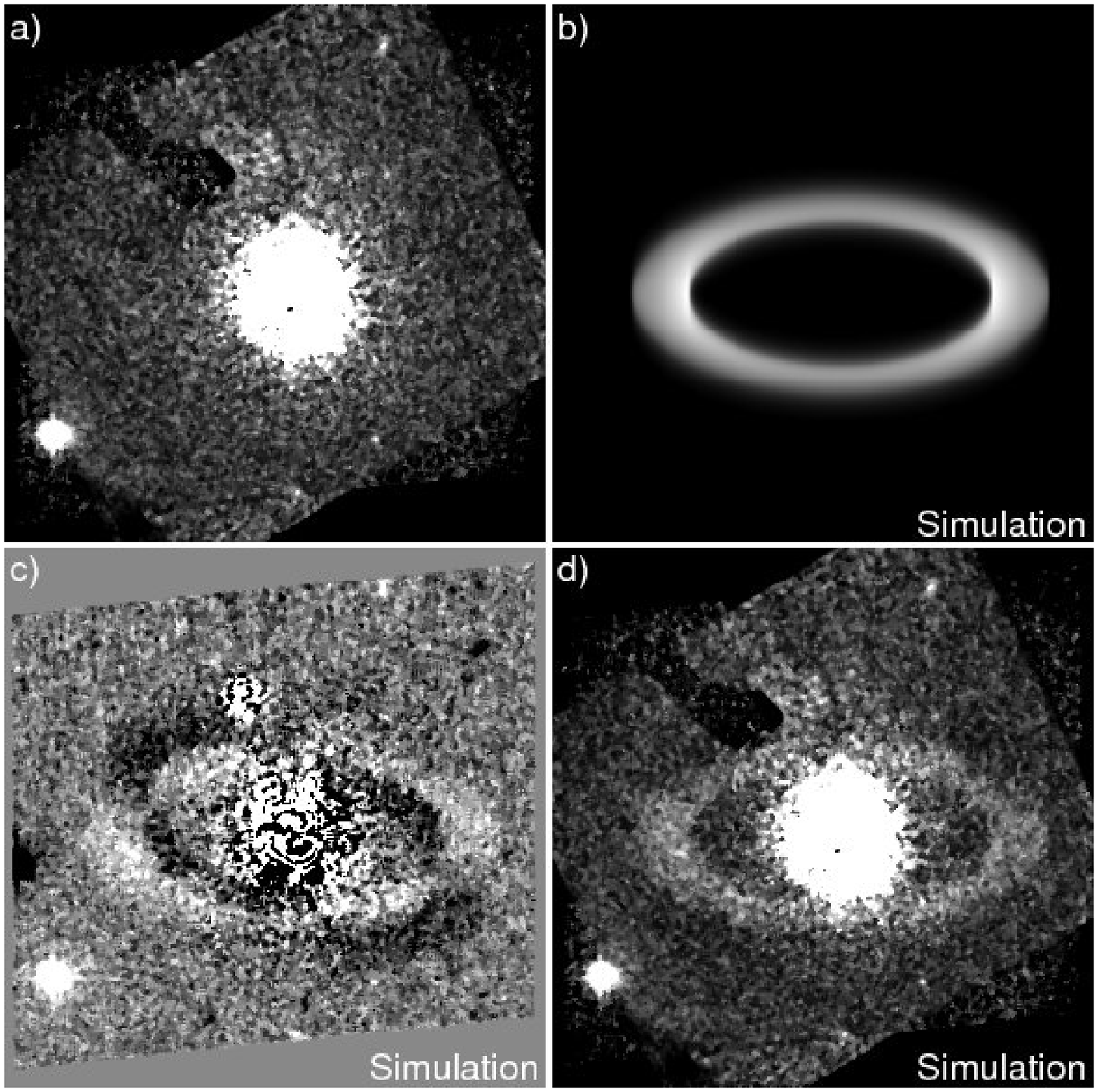} 
\caption{Experimental results of applying the roll subtraction algorithm to ACS
F606W coronagraphic images of HD 82943 taken at two orientations separated by
25\deg.  Each frame is 30\farcs 2 on a side.  (a) Result of the application of
the algorithm prior to adding a disk model to the images.  The residuals near
the center are due to PSF mismatches between the two orientations. (b) Model
disk that approximately matches the observed HD 207129 ring.  This is
added to each HD 82943 image (with a 25\deg orientation offset in the second
image) without noise and with twice the surface brightness of the HD 207129
ring.  (c) Direct subtraction of the image from the second orientation from
that from the first (with disk models added).  Positive and negative images of
the ring and background objects are seen.  (d) Application of the roll
subtraction algorithm to the ring-added images.  The resulting ring image
appears cleaner than that for HD 207129 because of the increased surface
brightness and lack of shot noise in the model.} 
\epsscale{1.0} 
\end{figure}

To verify that the subtraction algorithm can produce a reliable result, we have
tested our detection procedure on a simulated disk image combined with an
observed ACS coronagraphic PSF.  For the star we used ACS images of HD 82943,
which was also observed in our program with sequences similar to those for HD
207129.  A noiseless model of the HD 207129 ring with twice its surface
brightness relative to the star was added at each orientation and then
processed.  The results shown in Figure 2 indicate that this method can
reliably extract such a disk from these data.  Another experiment with the disk
model having a filled interior but otherwise the same inclination and size
showed that the extracted disk image also had a filled interior, so the inner
clearing in the real disk is not caused by self-subtraction.

It is unlikely that this disk would have been detected by subtracting a
reference star image, as we had originally intended with HD 211415.  There is
always a PSF mismatch due to color differences when a reference PSF is used.
This disk is so faint that it might have been detected only by using roll
subtraction and avoiding such color mismatches.  

\subsection{{\it Spitzer} Photometry and Spectroscopy}

The Multiband Imaging Photometer for {\it Spitzer} (MIPS) instrument was used
to image HD 207129 at 24 $\mu$m on 13 October 2004 as a part of GTO program 
41 (Beichman et al. 2005).  One cycle of the standard photometry dither pattern
was executed, producing a dataset with 16$\times$3s exposures.  Mosaics 
with 2\farcs 45 square pixels from the Spitzer Science Center (SSC) post-BCD 
pipeline version 14.4 (providing enhanced processing beyond the Basic 
Calibrated Data) were used for our analysis.  Photometry in a 
13\asec aperture yielded a total flux density of 164 $\pm$ 2 mJy, using the 
aperture corrections of Engelbracht et al. (2007).  This value is consistent 
with the one obtained by Trilling et al. (2008) from the same dataset.

MIPS 70 $\mu$m default scale imaging of HD 207129 was performed by Bryden et
al. (unpublished), and clearly showed the source to be extended.  Follow-up
imaging using the MIPS 70 $\mu$m fine scale channel was carried out on 03
November 2006 as a part of GO program 31057.  Five cycles of the small field
photometry dither pattern and 10 sec exposures were made at each of 4 cluster
target positions spaced on a square grid of 16\farcs 22 (3.16-pixel) spacing.
This observing strategy provides enhanced subpixel sampling to enable PSF
subtraction and deconvolution, and an effective exposure time of
5$\times$4$\times$84 = 1680 seconds on the source.  The data were processed
using the MIPS Data Analysis Tool (DAT; Gordon et al. 2005) version 3.10, with
final output mosaics made on a grid with 2\farcs 62 pixels.  The source appears
clearly extended, as expected from the prior default scale images.  Structure
in the sky background around the source complicates measurement of the source
flux.  Photometry of the fine scale data using a 15 pixel square aperture, with
an aperture correction derived from STinyTIM PSF simulations (Krist 2006),
finds 70 $\mu$m flux densities ranging from 244-317 mJy, depending on where the
background is chosen.  The midpoint of these values is consistent with the flux
density of 289 mJy reported for the default scale data by Trilling et al.
(2008), and is much brighter than the stellar photospheric emission.  However,
reprocessing of the default scale data using the most recent MIPS instrument
team pipeline produces images with a smoother background and fewer artifacts.
Photometry in those mosaics (4.925\asec/pix) using a 16 pixel aperture yields a
flux density of 369 $\pm$ 26 mJy.  As HD 207129 is a red source, a color
correction factor of 1.12 (assuming a color temperature of 50 K) must be
applied to yield a final 70 $\mu$m flux density of 413 $\pm$ 29 mJy.

MIPS 160 $\mu$m observations were carried out on 04 Jun 2007 as a part of GO
program 30157.  Three cycles of 10 sec exposures were made at each of nine
cluster target positions.  The target spacings provided offsets of 0\asec and
$\pm$36\asec (2.5 pixels) along the short axis of the 2$\times$20 pixel array,
and 0\asec, 72\asec, and 140\asec along the long axis to provide a larger area
for background measurements.  The total exposure time on the source was
$3\times9\times64$ = 1728 seconds.  Data processing was again with the MIPS DAT
ver 3.10, with the final product being a 130$\times$58 pixel mosaic at 4\asec
pixel scale.  The short-wavelength spectral leak is present adjacent to the
image of HD 207129, with intensity about half of the true source brightness.
This spurious source was removed by aligning and subtracting an image of $\tau$
Ceti, an object for which MIPS detects only the spectral leak at 160 $\mu$m.
Photometry of the leak-subtracted image is also complicated by variable 
background levels near the source.  Depending on the sky aperture placement, 
background-subtracted flux densities in a 16 pixel square aperture range 
from 218 to 283 mJy, with a 12\% overall calibration uncertainty.  We therefore
adopt a compromise value of 250 $\pm$ 40 mJy for the source at 160 $\mu$m.
This is somewhat larger than the value derived by Tanner et al. (2009) from 
a different MIPS dataset, and is about 30\% less than the value obtained from 
ISOPHOT observations by Jourdain de Muizon (1999) - probably because of source 
confusion in the large ISO beam (see Section \ref{sec:spit}).

MIPS Spectral Energy Distribution (SED) mode observations of HD 207129 were
performed on 27 Oct 2007 as a part of GTO program 30156.  This
observing mode produces R$\sim$15--25 spectroscopy over the wavelength range
from 55 to 90 $\mu$m.  The 120\asec$\times$20\asec slit was oriented along PA
229\deg, or roughly perpendicular to the disk major axis determined by Bryden
et al.  Fifteen cycles of 10 sec exposures were used, giving a total
integration time of 944 sec.  By chopping 1\amin off-source, a background
spectrum was obtained and subtracted from the source spectrum.  Reduction of
the spectral images was done using the MIPS DAT ver 3.10, which produced a
mosaic with 44 4\farcs 9 pixels in the spatial direction and 65 pixels in the
spectral direction.  This was then boxcar-smoothed over 5 adjacent rows to
improve the measured signal to noise ratio.  To extract the spectrum, the
signal in 5 columns centered on the peak emission was summed along each row of
the SED mosaic.  The resulting spectrum in instrumental units was converted to
physical units by reducing and extracting a MIPS spectrum of the calibration
star Canopus (F0 II) in the same way.  The Canopus spectrum was re-normalized
to give a flux density of 3.11 Jy at 70 $\mu$m, and corrected for the spectral
response function by assuming Canopus has a Rayleigh-Jeans spectral slope in
the far-IR.  The flux normalization and spectral response function derived from
Canopus were then applied to the MIPS spectrum of HD 207129, and the resultant
spectrum was further smoothed over adjacent wavelength bins.  This reduction
process assumes a point source slit loss correction; a revised slit loss
correction becomes necessary when fitting spatially extended source models to
these data (Section \ref{sec:model}).

{\it Spitzer's} InfraRed Spectrograph (IRS) was used to observe HD 207129 
on 10 June 2007 as part of GO program 20065.  Exposure times of
2$\times$6 sec, 3$\times$14 sec, and 5$\times$30 sec were employed for the SL1
(7-14 $\mu$m), LL2 (14-21 $\mu$m), and LL1 (20-28 $\mu$m) spectral modules
respectively.  The data were processed by the standard (SSC) pipeline version 
16.1.  Spectra were obtained for each order at two positions along the slit 
and extracted using the SSC SPICE (Spitzer IRS Custom Extraction) software 
package.  
The resulting flux is an average of these two nod positions, while the error 
bars are calculated from the difference between them.  The entrance slit 
for the SL spectrograph is just 3\farcs 7 wide, resulting in potential loss 
of some incoming flux depending on how well centered the target is within the 
slit.  The Long-Low slit width of 10\farcs 6 is much less prone to
pointing-related slit loss.  This is confirmed by comparison with the MIPS
photometry, where the IRS flux density at 24 $\mu$m (165 mJy) is in excellent
agreement with the aforementioned MIPS 24 $\mu$m value, and well within the
nominal $\sim$2\% calibration uncertainty for MIPS at 24$\mu$m (Engelbracht et
al. 2007).  Due to slit loss, however, the SL data had to be scaled
upward by a constant factor of 1.08 in order to properly overlap with the LL
data.  Following this scaling, data from the three spectral orders were spliced
together by simple removal of the low S/N edges of each order, resulting in a
final wavelength coverage from 7.6-35 $\mu$m.  Here also, a point source slit
loss correction is assumed and must be revised when modeling an extended source
(Section \ref{sec:model}).

\section{RESULTS}

\subsection{{\it HST} ACS}
\label{sec:HSTdata}

The ACS image reveals that the HD 207129 disk is a ring inclined
60\deg$\pm$3\deg from pole-on with its major axis along PA=127\deg$\pm$3\deg,
as measured by visually fitting ellipses.  The annulus is $\sim$1\farcs 9
($\sim$30 AU) wide with a mean radius of $\sim$10\farcs 2 (163 AU).  The ansae
have a surface brightness of $V$=$23.7\pm0.3$ mag arcsec$^{-2}$, which is
$22\times$ less than that of the unsubtracted coronagraphic stellar PSF at
those locations.  This makes the HD 207129 ring the faintest extrasolar 
circumstellar disk to have been imaged in scattered light (the Fomalhaut ring
surface brightness of $V$=22 mag arcsec$^{-2}$ makes it the faintest disk 
relative to the brightness of its star; Kalas et al. 2005).  The
similarity of the nebula surface brightness on opposite sides of the
ring major axis indicates a dust phase function with low scattering
anisotropy (see below), and thus provides no clear indication whether
the N or S side of the ring is foreground to the star.  Overall the ring 
structure appears azimuthally smooth; noise and subtraction residuals 
make it impossible to discern how uniform it is on small scales.  No strong 
constraint can be given on the presence of material interior to the ring: the 
shot noise remaining after PSF subtraction increases steeply inward 
(see Appendix), negating the effect of increased stellar illumination at 
smaller radial distances.

Three objects are detected near the ring.  A point source is seen 7\farcs 50
from HD 207129 at PA=196\deg, projected just outside the ring edge. It has a
brightness of $V$ = 22.8$\pm$ 0.3 (the uncertainty is largely due to the
non-uniform background of PSF subtraction residuals). A more diffuse source
that is likely to be a background galaxy is 5\farcs 6 away at PA=359\deg.
Another galaxy is 14\farcs 9 from the star at PA=115\deg.  An inspection of
Digital Sky Survey images shows that there are a moderate amount of both stars
and galaxies in the region, so lacking any color or proper motion information
to indicate otherwise, it is possible that the point source is a background
star and is not associated with HD 207129.  Further observations are needed to
test for companionship via common proper motion.

\subsection{{\it Spitzer} Imaging and Spectroscopy}
\label{sec:spit}

Mid and far-infrared images of HD 207129 and its surrounding field are shown in
Figure 3.  At 24 $\mu$m, the star appears as an unresolved point source.  To
check for faint extended emission, a 24 $\mu$m image of the reference star HD
217382 (which lacks any infrared excess) was aligned with that of HD 207129 and
subtracted.  The results show no significant extended emission from the disk at
24 $\mu$m.  

At 70 $\mu$m the source is clearly extended, with the fine scale source
well-fit by an elliptical Gaussian with FWHM 25\farcs 1$\times$17\farcs 7 
and major axis along PA 123\deg.  To retrieve the intrinsic source size from 
these values we must make comparisons to images of standard stars.  
The bright reference stars Altair, Sirius, and Procyon have a median 
elliptical Gaussian source size of 16\farcs 1$\times$ 15\farcs 3 in 
MIPS 70 $\mu$m fine scale images.  However, these sources lack IR excess, 
have a Rayleigh-Jeans slope across the 70 $\mu$m bandpass, and thus are 
significantly bluer than HD 207129.  They will thus appear slightly smaller 
than an unresolved source with strong excess, and this difference can be 
important to size estimates when the source extension is less than the 
telescope beamwidth.  By integrating the product of different spectral slopes 
with the MIPS 70 $\mu$m filter bandpass, we estimate that telescope 
diffraction should cause a flat spectrum source to appear 3.1\% larger 
than a naked photosphere.  Using this correction, quadrature subtraction 
of the beamsize from the data results in an intrinsic source size of 
18.8$\pm$0.8\asec (300$\pm$ 13 AU) along its major axis and 8.1$\pm$0.6\asec 
(130 $\pm$ 10 AU) along its minor axis.  The inclination and position 
angles implied for the ring are consistent with the $HST$ results, with the 
ring marginally smaller at 70 $\mu$m than seen in scattered light. 

At 160 $\mu$m the measured source size is 42\farcs 4$\times$34\farcs 0 
extended along PA 137\deg.  While this result is suggestive of a source
elongated at the same PA as seen at 70 $\mu$m and with $HST$, it must
be interpreted with caution.  Similarly-constructed 160 $\mu$m mosaics 
of other targets consistently show point sources instrumentally elongated 
by $\sim$20\% in the direction of MIPS scan mirror motion.  In the 
case of the HD 207129 160 $\mu$m observation, this instrumental axis is 
aligned within 20\deg of the observed source extension.  Furthermore, 
imperfections in subtraction of the 160 $\mu$m spectral leak could
create systematic errors in the source profile that will confuse our 
ability to measure small deviations from an unresolved source.  An intrinsic 
FWHM of the 160 $\mu$m source thus cannot be reliably determined, and we 
can only exclude source diameters larger than about half the beamsize 
(20\asec / 320 AU).

\begin{figure}
\plotone{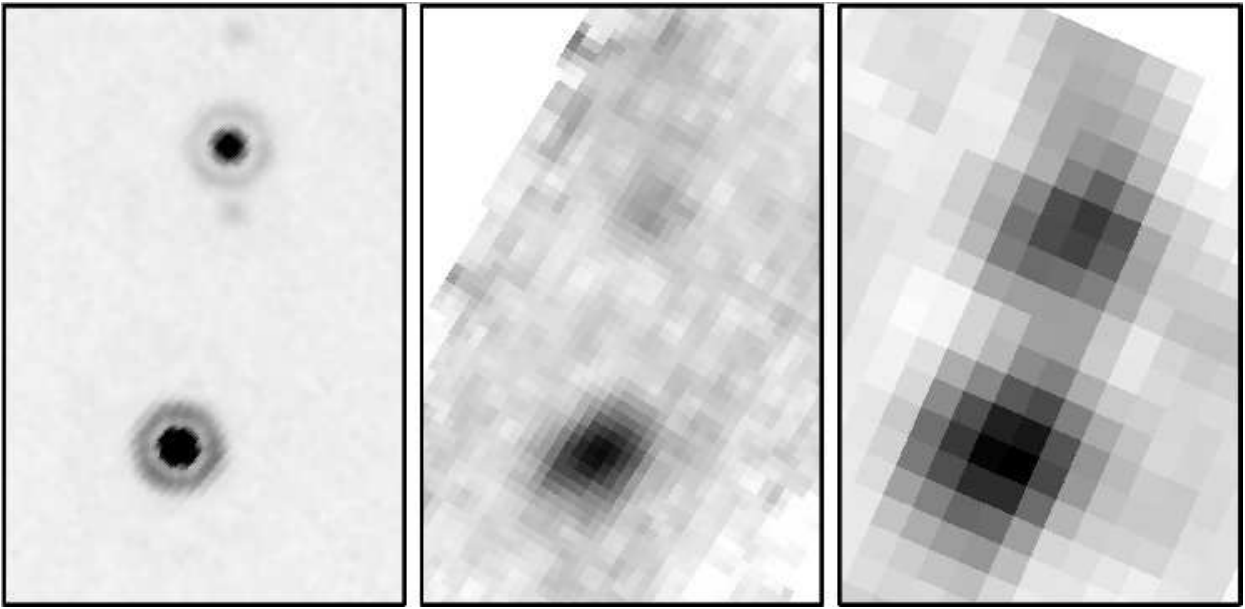}
\caption{{\it Spitzer} MIPS images of HD 207129 (lower center), from left
to right at 24, 70, and 160 $\mu$m respectively.  The field of view is
100\asec$\times$150\asec, with N up and E to the left.  The 24 $\mu$m
image is shown in log stretch, while the 70 and 160 $\mu$m images are in 
linear stretch.  Only the 70 $\mu$m fine scale image is shown.  The spectral 
leak has been subtracted from the 160 $\mu$m image shown.  The star 
CD-47 13929 is seen 75\asec N of HD 207129 at 24 $\mu$m.  A very red 
background object is located 59\asec N of HD 207129.}
\end{figure}

The spectral energy distribution of HD 207129 is plotted in Figure 4.  Excess
emission first becomes evident near 28 $\mu$m, rises steeply to a plateau
between 60-90 $\mu$m, and then falls off toward longer wavelengths.  The wavy
pattern in the MIPS SED spectrum between 50 and 90 $\mu$m is an artifact of the
spectral extraction process.  At 160 $\mu$m, MIPS measures a flux density only
about half the ISO values reported by Jourdain de Muizon (1999).  This is
likely due to source confusion in the large-beam ISO measurements, which
included both the star and the (presumably extragalactic background) source
about 1\amin to the north.  The new 870 $\mu$m continuum detection of the 
source by Nilsson et al. (2010) is shown.  At 24 $\mu$m, the measured flux 
density is comparable to the photospheric emission value of 160 mJy derived 
from a Rayleigh-Jeans extrapolation of the IRS SL measurements.  Trilling et 
al. (2008) had reported a 24 $\mu$m excess for HD 207129, based on a 
photospheric flux density estimate 15\% smaller than our value.  It now 
appears that Trilling's photosphere flux density estimate was adversely 
affected by saturated 2MASS photometry, and should be superseded by the value 
given here from the IRS SL spectrum.  Thus the star has no excess emission 
at 24 $\mu$m.
 
\begin{figure}
\includegraphics[scale=0.7]{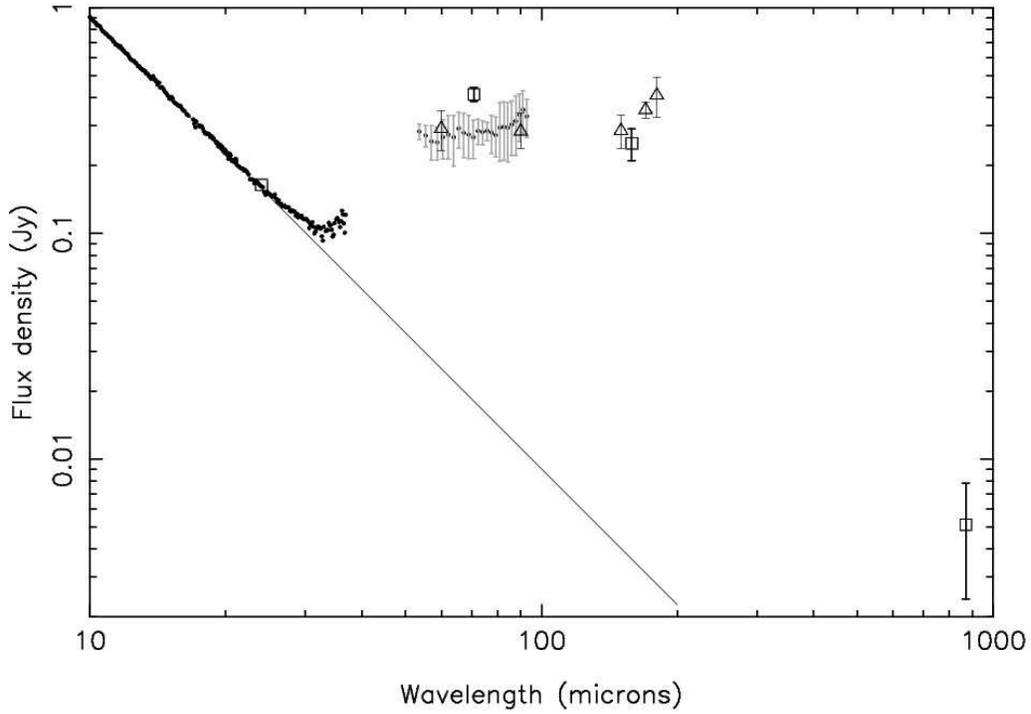}
\caption{Spectral Energy Distribution of HD 207129.  Individual data
points show the IRS and MIPS SED results, while the three MIPS photometry
points are plotted as open squares.  ISO measurements appear as open 
triangles.  The solid line shows a Rayleigh-Jeans extrapolation of the 
stellar photosphere emission based on the IRS data near $\lambda=$ 10 $\mu$m. 
The MIPS-SED spectrum shown here includes a slit loss correction for a point
source; therefore, it is low compared to the 70 $\mu$m photometry
which includes extended emission.  The 870 micron point is from 
Nilsson et al. (2010).
}
\end{figure}

\section{A Combined Image and SED Model}
\label{sec:model}

The detection of the ring in scattered light provides a new
opportunity to understand the overall properties of this debris system.
Previous work based only on the spectral energy distribution inferred
a large disk outer radius of 500-1000 AU, large disk inner hole of 
200 AU, and particle sizes ranging from 1-200 $\mu$m (Jordain de Muizon 
et al. 1999).  Our $HST$ and {\it Spitzer} images show a much smaller 
and narrower ring.  Given this new information on the spatial scale of 
the system, and the improved far-infrared photometry and spectroscopy
provided by {\it Spitzer}, an updated model analysis of the disk is now
called for.

We begin with the simple assumption that the disk emission originates in a
single radial zone specified by inner and outer radius, a radial power law
surface density, Gaussian vertical scale height and radial flaring exponent,
and astronomical silicate grains whose wavelength-dependent emissivities are
calculated from Mie theory with the optical constants of Laor \& Draine 
(1993).  Dust thermal emission is calculated assuming local
thermal equilibrium with a 1.2 \Lsun star (Bryden et al. 2006) for each of the
grain sizes considered in the distribution between minimum and maximum radii
a$_{min}$ and a$_{max}$.  The grain size distribution is at first assumed 
to follow an a$^{-3.5}$ power law appropriate to a collisional cascade 
(Dohnanyi 1969), but other slopes are considered.  Initial values for the 
ring geometrical parameters are taken from the $HST$ image (section 
\ref{sec:HSTdata}).  Model thermal images are then calculated for this dust 
distribution on a 1\asec spatial grid for 31 wavelengths spanning 10-850 
$\mu$m.  The total flux in these 31 channels is compared to the SED of the 
infrared excess emission, while appropriate subsets of model images are 
combined to synthesize broad-band images as observed by the MIPS 24, 70, 
and 160 $\mu$m cameras.  Model images within the 7--35 $\mu$m region sampled 
by the IRS data, and within the 55--90 $\mu$m region sampled by the MIPS SED 
mode, are convolved with the instrumental point spread function and windowed 
by synthetic entrance slits appropriate to the instrument and slit position 
angle used in the {\it Spitzer} observation.  A model scattered light image 
is calculated for the same dust density distribution on a spatial grid of 
1\asec, with the dust albedo and phase function asymmetry parameter as 
additional inputs.  Isotropic scattering is assumed.

Comparison of model spectra to the observations requires careful
consideration of slit losses.  The IRS and MIPS SED data are 
normally calibrated with slit loss corrections appropriate to a point 
source, with the goal of making their flux calibration consistent with 
large-aperture MIPS 24 and 70 $\mu$m photometry.  For a spatially extended 
source such as the ring of HD 207129, a point-source slit loss correction 
is no longer appropriate for calibration of these spectra.  The true slit
loss correction will depend in detail on the specific source brightness 
distribution, which is model-dependent.  Rather than derive slit loss
corrections and recalibrate the spectral data for each possible model
SED, we choose to compare model spectra that have suffered uncorrected
slit losses to a version of the {\it Spitzer} data that has had its point-source 
slit loss correction removed.  Slit loss correction factors for the IRS SL 
and LL data were derived using the IRS CUBISM software v1.6 (Smith et al. 
2007).  MIPS SED slit loss correction factors were calculated by windowing 
STinyTIM PSFs (Krist 2006) through synthetic entrance slits at six wavelengths 
spanning 50--90 $\mu$m and interpolating as needed for intermediate 
wavelengths.  A nominal slit width of 20\asec was assumed. 

We seek a global model that can reproduce the infrared photometry, spectroscopy,
resolved source size at 70 $\mu$m, and the scattered light brightness
seen in the $HST$ image.  With its superior spatial resolution, the ACS
coronagraphic image fixes the radial location of the dust within the
148--178 AU region, as well as the ring inclination and position angle.
An optimal model for the {\it Spitzer} data was sought manually by varying 
the normal optical depth and the minimum/maximum silicate grain radii
a$_{min}$ and a$_{max}$.  Our initial guess for a$_{min}$ was 0.6 $\mu$m, 
the geometric blowout size for grains with density 2.5 gm cm$^{-3}$, 
stellar luminosity of 1.2 \Lsun, and stellar mass of 1.1 \Msun 
(equation A19 of Plavchan et al. 2009; values from 
http://nsted.ipac.caltech.edu).

An initial solution was found for model parameters with a$_{min}$ of 1.4
$\mu$m, a$_{max}$ of 400 $\mu$m, and a normal geometrical optical depth of
0.013 for grains with radius a$_{min}$ at r$=$ 163 AU.  The model reproduces
the observed FWHM of the source at 70 $\mu$m; the slightly smaller source
diameter at 70 $\mu$m, relative to the ring seen in scattered light, can 
be understood by the warmest emission being concentrated at the ring inner 
edge.  Integrating over the grain size
distribution, and assuming a mean grain density of 2.5 gm cm$^{-3}$, the total
dust surface density is 4$\times10^{-6}$ gm cm$^{-2}$ at r$=$ 160 AU. and the
dust mass in the model integrated over the full spatial extent of the ring is
0.07 lunar masses.  A sharp ring inner edge provides a good match to the
steeply rising excess emission beyond 30 $\mu$m, and the lack of scattered
light emission inside radii of 148 AU.  We found that the nominal -3.5 power
law slope caused the model to over-predict the 160 $\mu$m flux density relative
to 70 $\mu$m.  By steepening the grain size distribution to a -3.9 power law,
an adequate fit is obtained.  This change strongly suppresses the submillimeter
flux from the model; to compensate, a larger value of a$_{max}=$ 700 $\mu$m is
adopted.  A slope $\le$$-$3.7 is compatible with the
measurements within the errors.  This slope is preferred by recent numerical
simulations of the equilibrium grain size distribution when material strength
effects are considered (A. G\'asp\'ar, in preparation).

\begin{figure}
\includegraphics[scale=0.7]{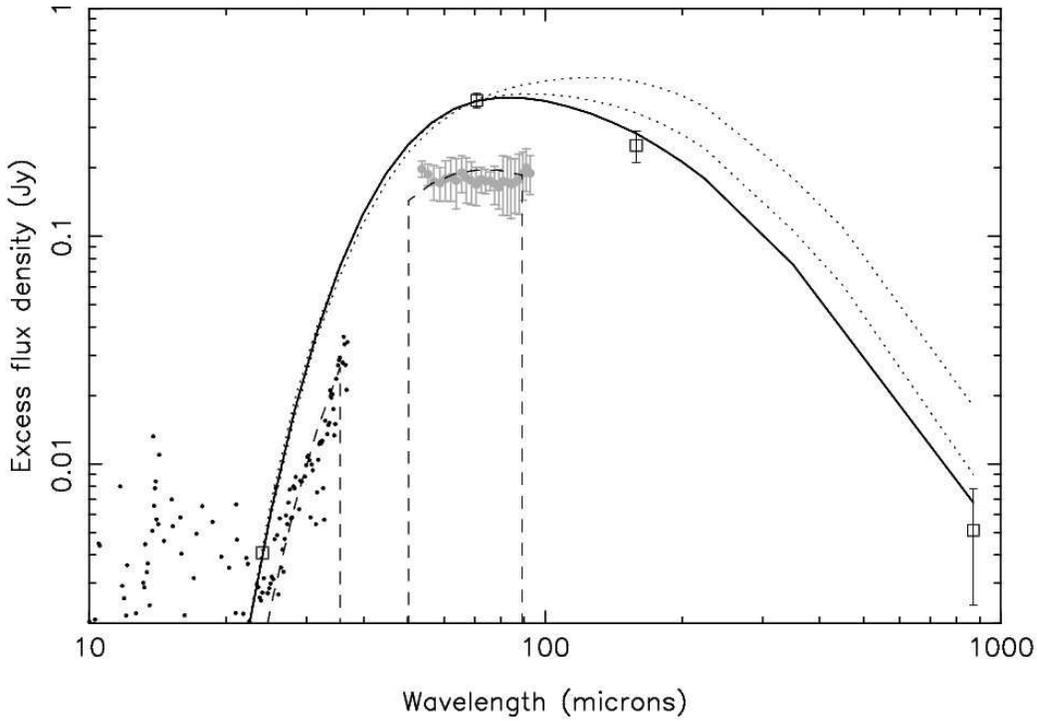}
\caption{Infrared excess of HD 207129 (points), compared to our preferred
disk model (solid line) and the disk model windowed by the IRS and MIPS SED
slits (dashed lines).  The stellar photosphere emission has been removed
using a Rayleigh-Jeans slope normalized to the IRS SL data.  Relative to
Fig. 4, the excess flux densities measured in the IRS and MIPS SED spectra 
have been reprocessed to remove the point source slit loss correction. 
The effects of changing the dust size distribution slope from -3.5 
(upper dotted curve) to -3.7 (lower dotted curve) to -3.9 (solid
curve) are shown. All of the models have the modified albedo.}

\end{figure}

While this model provides a good fit to the available photometry and the
spectrophotometry measured through the IRS and MIPS SED slits, it fails 
in its predicted scattered light brightness.  According to Mie theory, 
astronomical silicate grains larger than 1 $\mu$m radius should have an 
albedo $\omega=$ Q$_{scat}$/(Q$_{abs}$ + Q$_{scat}$) of $\sim$55\% in the 
optical and near-infrared.  This value was assumed in the thermal equilibrium 
calculations intrinsic to the SED model.  For this value, and integrating over 
the grain size distribution of our preferred model, the model ring is much 
brighter in scattered light than the $HST$-measured surface brightness on 
the ring ansae.
 
The fraction of the stellar luminosity scattered by the grains, relative 
to the total luminosity incident on the grains, provides a wavelength-averaged 
albedo that can be estimated from the available data:

$$ <\omega> = F_{scat} / (F_{scat} + F_{emit}) $$

\noindent
where $F_{scat}$ is the fractional scattered light luminosity of the
disk (relative to direct starlight), and $F_{emit}$ is the fractional
infrared luminosity of the disk = 1.4$\times10^{-4}$ from the {\it Spitzer}
data.  $F_{scat}$ is estimated from the $HST$ images by renormalizing 
our scattered light model image of the ring so that its surface brightness
on the ansa matches the observed value, and then adding up the total
scattered light in the model ring structure.  The result finds that 
$F_{scat}\approx$ 7.6$\times10^{-6}$.  To the extent that this value,
measured at 0.6 $\mu$m, is representative of the fraction of scattered to
direct starlight averaged over the wavelengths that produce significant 
stellar heating of the grains, the mean albedo of the dust particles 
would be $<\omega>\approx 5.1\%$.  Simple Mie theory grains are thus not 
consistent with the combined suite of observations for HD 207129. 

When the SED modeling is repeated using this reduced albedo, but
retaining the original Mie values for Q$_{scat}$, a very similar model
is obtained where the only required change is that $a_{min}$ increases
to 2.8 $\mu$m.  The higher grain emissivity causes larger grains to
come to the same equilibrium temperature as the smaller, more reflective
grains initially assumed.  In making this adjustment, we have retained
the emissive properties of astronomical silicate grains but modified
their reflectivity: essentially painting the grains black.  Porous
dust grains might account for this results, as they are thought to
have lower albedoes (Hage and Greenberg 1990) and larger blowout sizes 
(Saija et al. 2003) in comparison to solid grains made from the
same material.

The best matching SED model with the modified grain albedo is shown 
in Figure 5.  Future work will be needed to assess what combination of 
grain properties and scattering theory can provide a self-consistent 
solution to the emissivity and albedo values we found necessary to fit 
the optical and infrared properties of the HD 207129 ring.

\section{Discussion}

HD 207129 joins the small but growing list of stars that have resolved debris
disks imaged in scattered light, which as of the time of writing number nearly
20.\footnote{See http://circumstellardisks.org for an up to date census of
resolved disks.} The ring-like appearance of its disk is similar to those of HR
4796 and Fomalhaut, whose central cleared zones suggest the presence of low
mass substellar companions that tidally remove dust. The most prominent example
of this behavior is the Fomalhaut ring (Quillen 2006; Chiang et al. 2009),
which is eccentric, has a sharp inner edge, and for which the predicted
companion has been subsequently imaged (Kalas et al. 2008).  The low definition
of the HD 207129 ring makes it difficult to measure its properties precisely.
A sharp ring inner edge is indicated by the steeply rising excess emission
beyond 30 $\mu$m, and is consistent with the absence of scattered light inside
a radius of 150 AU.  A planet sculpting the ring may lie just inside this
radius, but no field objects are seen there in our coronagraphic images.  
The point source projected near the S outer edge of the ring could be
relevant to the ring dynamics, if it was found to be a co-moving member
of the system.  Its brightness in F606W would correspond to a >20 
M$_{Jupiter}$ brown dwarf, according to the spectral evolutionary model 
  of Burrows, Sudarsky, and Lunine (2003).

To constrain a possible offset of the ring center from the star, we visually
overlaid ellipses on the ring image (Figure 6) to fit for the ring center.
We find that any offset of the star from the ring center must be smaller than 
0.4\asec and 0.2\asec along the ring minor and major axes (respectively).  
The corresponding upper limit to any ring eccentricity is 0.08, for the case 
where the line of apsides would be projected farthest from the plane of the 
sky.  

\begin{figure} 
\includegraphics[scale=0.5]{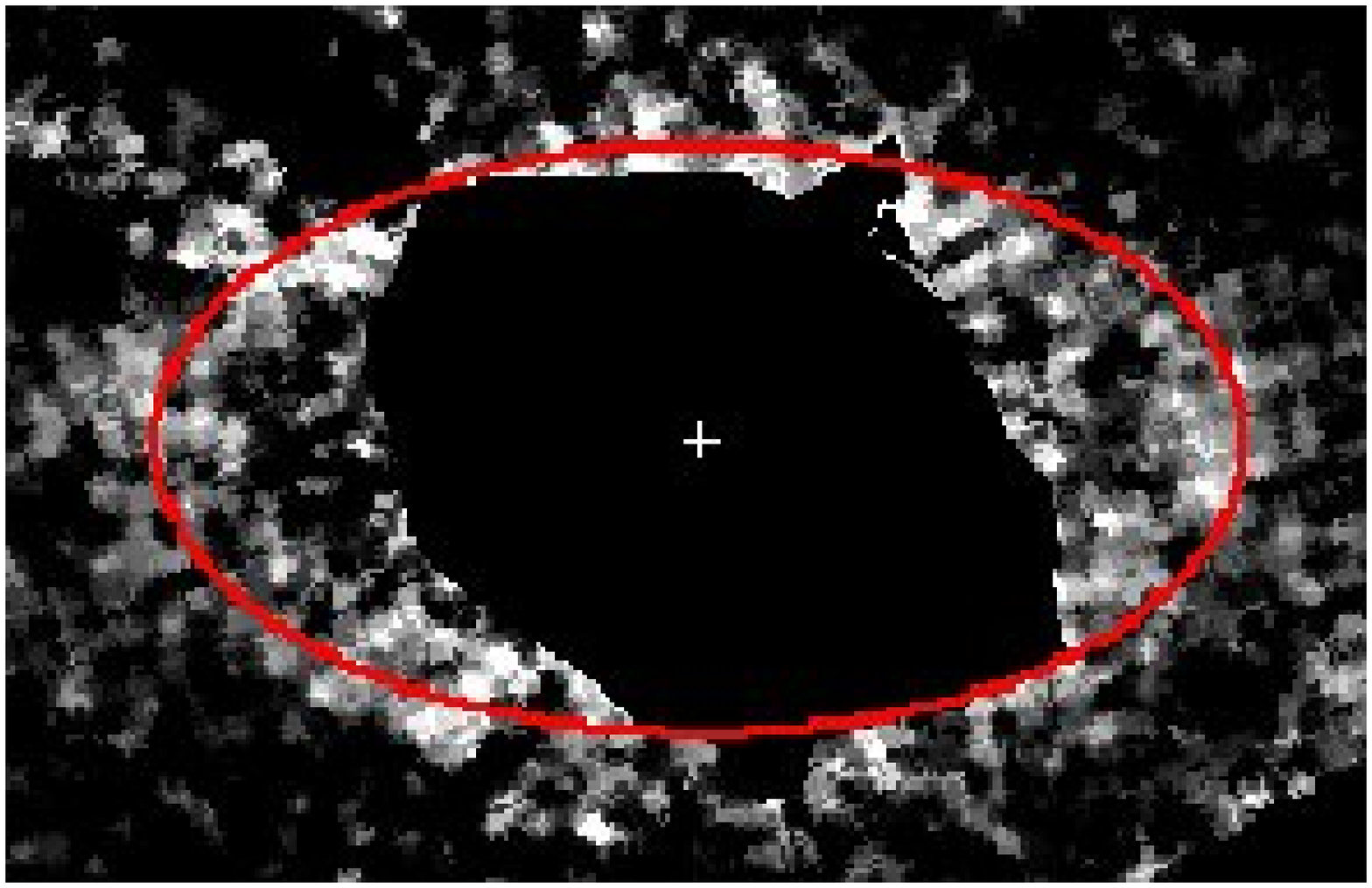}
\caption{The HD 207129 ring ACS image with an ellipse superposed centered on 
the star (marked by the cross) and having a semi-major axis of 10''.  No large 
offset of the ring relative to the star, like that seen for Fomalhaut, is seen 
that would suggest a significant planetary perturber.} 
\label{fig:offset}
\end{figure}

The low signal of the HD 207129 ring in the ACS images limits our measurement
accuracy and thus only rough characteristics can be derived.  The lack of a
significant difference in brightness in the near and far sides of the ring
indicates nearly-isotropic scattering.  Based on comparisons to our approximate
ring models (Figure 7), we constrain the asymmetry parameter, $g$, of the
Henyey-Greenstein scattering phase function to be $<0.1$.  This low level of
forward scattering is only matched by the disk of HD 92945 (Golimowski et al.,
in prep).  Other debris disks, like Fomalhaut (Kalas et al. 2005), HD 141569a
(Clampin et al. 2003), AU Mic (Krist et al. 2005), and HD 107146 (Ardila et al.
2004) have more forward scattering ($g>0.15$). 

\begin{figure} \includegraphics[scale=0.7]{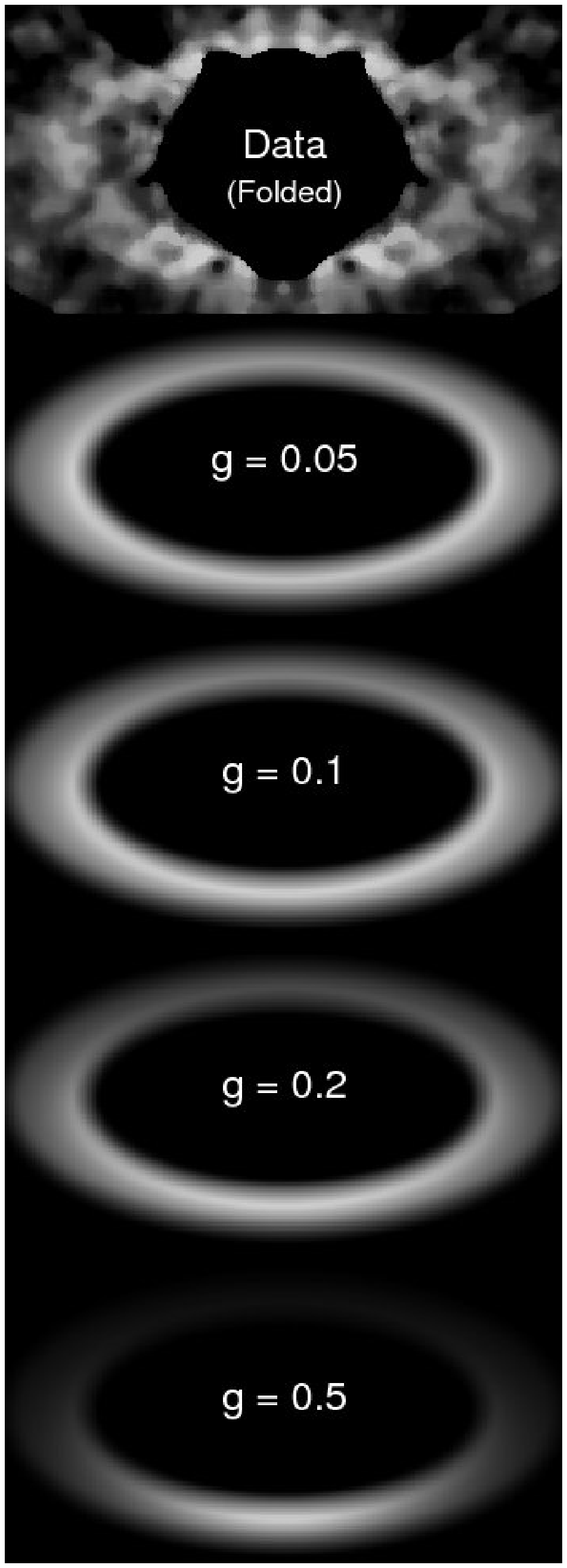} 
\caption{
At the top is the ACS image of the HD 207129 ring. To improve the signal, a
copy of the original image was flipped about the star (left semi-major axis
onto right semi-major axis), the two added together, and then smoothed.  Below
it are three-dimensional scattered light disk models with different values of
the Henyey-Greenstein phase term, $g$. These demonstrate by comparison that the actual ring
does not have significantly forward scattering (large $g$) grains. All images
are displayed with linear intensity stretches between their minimum and maximum
values. }  
\label{fig:g} 
\end{figure}
 
The nearly isotropic scattering in the ring conflicts with any assumption of
spherical particles that scatter according to Mie theory.  Our combined
scattering+SED modeling indicates that the minimum grain size is $\sim$2.8
$\mu$m.  Spherical 2.8$\mu$m particles are predicted to be strongly forward
scattering for wavelengths less than the grain size ($g\approx$ 0.8 at
$\lambda= 0.6 \mu$m).  If such grains were present in the HD 207129 ring, then
the foreground portion of the ring should appear much brighter than the ansae
or back side.  This effect is not seen in the $HST$ images.  In Mie theory,
astronomical silicate grains with characteristic radius $<$ 0.05 $\mu$m would be
needed to match this result, and would appear much warmer than the $\sim$ 44 K
temperature that characterizes the spectral energy distribution.  Combined with
the low derived albedo, this result points to non-spherical, possibly porous or
coated grains.  

Previous estimates for the age of HD 207129 have ranged from as little as
30-40 Myrs (Zuckerman, Song, \& Webb 2001) to as high as 4.4 -- 8.3 Gyrs
(Lachaume et al. 1999; Trilling et al. 2008).  The very young age estimate
has been withdrawn, but an age as young as 600 Myrs is still suggested
(Song, Zuckerman, \& Bessell 2003,2004).  A recent recalibration of the
dependence of chromospheric activity on stellar age suggests an age of 2.1
Gyrs (Mamajek \& Hillenbrand 2008) based on its Ca II emission strength
of log $R^{\prime}_{\rm HK}$= -4.8 (Henry et al. 1996).  This is consistent
with an age estimated based on Li abundances measured by Soderblom (1985).
However, the X-ray luminosity of this source
(1.5$\times10^{29}$ erg sec$^{-1}$, based on the ROSAT source counts and a
nominal spectrum) is considerably larger than typical for members of the
NGC 752 cluster at age 1.9 Gyrs (Giardino et al. 2008).  An age of 1 Gyr
is plausible.

The detection of this disk in scattered light is rather fortuitous.  As
measured by its 10$^{-4}$ fractional emission luminosity, it is among the
lowest dust content disks to have been imaged.  If the material were more widely
distributed radially around the star (such as in the case of $\beta$ Pic)
instead of concentrated in a ring, its surface brightness would be
significantly reduced.  In addition, if the star were 30 pc away rather than
$\sim$15 pc, the disk would appear half as far from the star and would be lost
in the greater glare and PSF subtraction residuals that exist there.  The star
is also bright enough ($V$=5.6) for sufficient scattered light to be detected.
These same conditions apply for Fomalhaut ($d$=7.7 pc, $V$=1.2), which has a
somewhat lower fractional excess ($\sim8\times10^{-5}$).  These examples
highlight the considerations required when optimizing a target list for
coronagraphic observations, especially given the limited observing resources on
$HST$ in both time and coronagraphic performance.  Targets that are too far
away and have too low an excess are poor candidates with little chance of a
detection. 

\section{Conclusions}

$HST$ coronagraphic images detect the debris disk of HD 207129 as a narrow 
ring with radius 160 AU, significantly smaller than previous estimates based
only on analysis of the source SED.  The ring size and orientation are
comparable to that inferred from resolved MIPS images of the
source at 70 $\mu$m.  This ring is among the faintest circumstellar 
features ever detected with $HST$, and was only reliably extracted from
stellar PSF artifacts through the use of a roll self-subtraction technique.

Given the new definition of the system geometry, we fit a model to the {\it Spitzer}
images and spectrophotometry.  We find that the observed emission is consistent
with material in the single radial zone where $HST$ detects scattered light, and
with dust grains ranging from 2.8 $\mu$m to at least 500 $\mu$m radius.
However, the almost isotropic scattering properties and low albedo of the ring
particles are not consistent with simple Mie theory estimates for silicate
grain composition.  The narrowness of the ring, its apparently sharp inner
edge, and large central cleared region are similar to the Fomalhaut system, and
suggest that deep near-IR imaging searches for substellar/planetary companions
might be profitable in the HD 207129 system.  Further observations with {\it
Herschel} could refine the ring emission properties for a more detailed 
comparison to the $HST$ images, while far-IR spectroscopy might provide some 
better indication of the dust composition.

\acknowledgments {
We thank Paul Smith (U. of Arizona) for his assistance with the MIPS SED data,
and Karl Misselt and Viktor Zubko for calculating optical properties of grains
at large size parameters.
This work was supported by Hubble Space Telescope General 
Observer Grant 10539 to the Jet Propulsion Laboratory, California Institute 
of Technology and by the {\it Spitzer} Project Science Office at JPL.  
Funding from both was provided by the National Aeronautics and Space 
Administration.  The {\it Spitzer Space Telescope} is operated by the 
Jet Propulsion Laboratory, California Institute of Technology, under 
NASA contract 1407.  
}

\section{Appendix}

In addition to HD 207129, we observed seven other debris disks selected from
{\it Spitzer} results to have fractional infrared luminosities $\ge10^{-4}$.
The disk of HD 10647 (distance $=$ 17 pc) was detected and will be reported in
Stapelfeldt et al. (in prep).  The other six are listed in Table
\ref{tab:targs}.  None of these showed scattered light at the contrast and
inner angle limit ($r \approx$ 1\farcs 0) of the ACS coronagraph.  The
non-detections provide new constraints on the disk properties as these
observations represent the highest contrast observations to date of those
targets.

All of the candidates were observed using sequences similar to those used for
HD 207129, with small variations in exposure lengths and roll separations
(20\deg -- 30\deg) as imposed by {\it HST} pointing constraints.  Reference
stars (specified in Table \ref{tab:targs}) were observed and their images
subtracted as described for HD 207129.  Each subtracted image contained a
faint, symmetric halo around the star caused by PSF mismatches that were
largely due to color differences but also time-dependent changes in the optical
system.  These halos could hide potential, otherwise-detectable face-on disks.
No disks were detected using reference PSF subtractions.  The iterative roll
subtraction method was also applied to these targets, using $4\times4$ rebinned
data, and no disks were detected. 

To estimate extended-source detection limits in the roll-subtracted images,
1\asec$\times$1\asec uniform-intensity squares were added to the two
unsubtracted images of HD 82943 at various locations with appropriate
orientations.  These were then $4\times4$ rebinned and processed with the roll
subtraction algorithm.  This process was continually repeated with adjustments
made to the squares' brightnesses until they appeared at their minimum robust
visual detection limit.  The local root-mean-squared (RMS) noise was measured
in the same locations as the squares but in a square-free roll-subtraction.  It
appeared that the detection limit (in flux per pixel) for the squares was
approximately equal to $1.2\times$ the local RMS per-pixel noise.  

For each roll-subtracted image of each target, the mean azimuthal RMS was
computed at each radius centered on the star in the subtracted image.  In this
calculation, care was taken to avoid background objects such as galaxies and
field stars.  These RMS noise curves vs. radius were then converted to
detection limits using the factor of 1.2 and were then converted to $V$
magnitude surface brightnesses.  The resulting curves in Figure 8 show that
read noise and perhaps sky background noise at large angles set the detection
limit to about 24 mag arcsec$^{-2}$ regardless of star brightness.  The
detection limit increases closer to the star where subtraction residuals, which
scale with star brightness, dominate. 

While the detection limits presented here are expressed in absolute units, what
really matters is the limiting brightness relative to the star, given that a
debris disk is optically thin and scales in brightness with its star.  A
relative detection limit can be converted to a maximum allowed optical depth
along the line of sight.  This can then be used to place constraints on some
disk properties, but only if a number other properties are known as multiple
parameters can affect line-of-sight optical depth: dust density, albedo,
disk height, inclination, degree of forward scattering, etc.  The maximum
allowed optical depth ($\tau$) times the albedo ($\omega$) at a given radius
$r$ (arcseconds) from the star is $\omega \tau = 4 \pi r^2 \times 10^{-0.4(SB -
V_{star})}$, where $SB$ is the detection limit in mag arcsec$^{-2}$ and
$V_{star}$ is the magnitude of the star.  Note that $\omega$ includes any
possible brightening of one side of a disk due to asymmetric scattering.  In
general, $\omega \tau$ at 3\asec is about 1 -- 2 $\times 10^{-4}$ for these stars.

Possible reasons for not detecting a disk around one of these six candidates
include a low grain albedo (which would render the disk faint), a small disk
angular radius (which could leave the system hidden behind the occulting spot),
or a face-on orientation (which would make the disk more difficult to
distinguish from PSF residuals). 

\begin{figure} 
\plotone{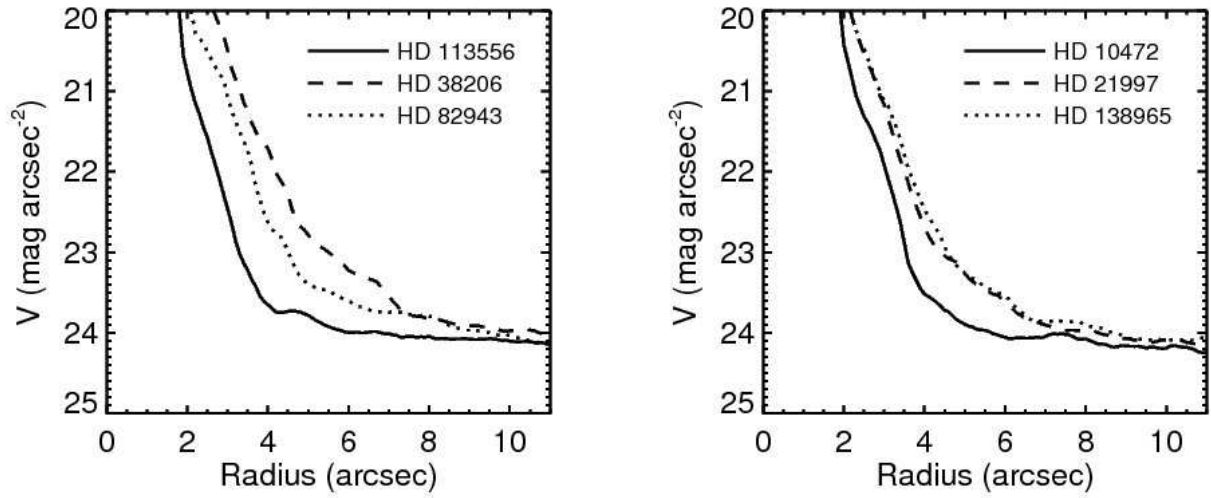} 
\caption{Lower limits for the reliable visual detection of a 1''$\times$1''
uniform intensity square around six stars for which no disks were detected.} 
\label{fig:uplims}
\end{figure}

Among the targets surveyed, HD 38206's low $v\sin{i}$ (Royer et al. 2002) may
indicate that the disk is projected nearly face-on (assuming the disk is
coplanar with the star's equator) and thus has a low line-of-sight optical
depth.  HD 82943, the nearest of the six disk non-detections, has two radial
velocity planets near 1 AU (Mayor et al. 2004).  Ground-based adaptive optics
imaging also failed to detect this disk in scattered light (Sch\"utz et al.
2004); due the highly stable PSF of $HST$, our limiting surface brightness
(Fig. \ref{fig:uplims}) is 5 mag better.  This system is comparable to 
HD 207129 in terms of its stellar type, V magnitude, fractional disk 
luminosity, and infrared excess SED.  A ring comparable to that of HD 207129
would have an apparent radius of 6\asec in the HD 82943 system.  Scattered
light residuals at this radius are only a bit worse than at 10\asec
where the HD 207129 ring was detected, so the non-detection of HD 82943's
disk in scattered light indicates different grain properties, a more face-on 
orientation, or a greater radial width of the emitting region.  The disk of 
the most distant candidate, HD 113556, would go undetected if its outer radius 
was less than 150 AU.  Based on the observed excess ratio between 24 and 70 
$\mu$m, all eight disks we studied have a characteristic blackbody temperature 
between 60 and 80 K.  

The most prominent difference between the detected and undetected disks is 
their proximity: $d<$20 pc for the two detected systems.  This suggests that 
a disk's angular extent may be the most important property for detectability 
with $HST$, both to avoid blockage by the occulting spot and the higher PSF 
subtraction residuals found closer to the star.  Future detection surveys 
for debris disks in scattered light should focus on the nearest (suitably 
bright) stars with strong infrared excess.
 
\begin{deluxetable}{lcrrcccl}
\rotate
\label{tab:targs}
\tablecolumns{8}
\tablewidth{0pc}
\tablecaption{Debris Disk Non-Detections in {\it HST} Program 10539}
\tablehead{
\colhead{} & \colhead{Spectral} & \colhead{$V$} & \colhead{Distance} &
\colhead{} & \colhead{Observation} & \colhead{$L_{\rm dust}/L_{\star}$} & 
\colhead{{\it Spitzer}} \\
\colhead{Target} & \colhead{Type} & \colhead{(mag)} & \colhead{(pc)} & 
\colhead{PSF Star} & \colhead{Date} & \colhead{$\times10^{-4}$} &
\colhead{Reference}}
\startdata
HD  10472 & F2IV/V & 7.6 &  67 & HD  12894 & 02 Oct 2005 & 7 & Rebull et al. (2008)\\
HD  21997 & A3IV/V & 6.4 &  74 & HD  15427 & 12 Aug 2006 & 5 & Mo{\'o}r et al. (2006)\\
HD  38206 & A0V    & 5.7 &  69 & HD  41695 & 01 Nov 2005 & 1 & Su et al. (2006)\\
HD  82943 & G0     & 6.5 &  27 & HD  84117 & 23 Nov 2005 & 1 & Trilling et al. (2008)\\
HD 113556 & F2V    & 8.2 & 102 & HD 101727 & 13 Apr 2006 & 7 & Chen et al. (2005)\\
HD 138965 & A1V    & 6.4 &  77 & HD 167468 & 01 Jun 2006 & 4 & Morales et al. (2009)\\
\enddata
\end{deluxetable}

\begin{thebibliography}

\bibitem[Ardila et al.(2004)]{2004ApJ...617L.147A} Ardila, D.~R., et al.\ 
2004, \apjl, 617, L147 

\bibitem[Beichman et al.(2005)]{2005ApJ...622.1160B} Beichman, C.~A., et 
al.\ 2005, \apj, 622, 1160 

\bibitem[Bryden et al.(2006)]{2006ApJ...636.1098B} Bryden, G., et al.\ 
2006, \apj, 636, 1098 

\bibitem[Burrows et al.(2003)]{2003ApJ...596..587B} Burrows, A., Sudarsky, 
D., \& Lunine, J.~I.\ 2003, \apj, 596, 587 

\bibitem[Chen et al.(2005)]{2005ApJ...623..493C} Chen, C.~H., Jura, M., 
Gordon, K.~D., \& Blaylock, M.\ 2005, \apj, 623, 493 

\bibitem[Chiang et al.(2009)]{2009ApJ...693..734C} Chiang, E., Kite,
E., Kalas, P., Graham, J.~R., \& Clampin, M.\ 2009, \apj, 693, 734

\bibitem[Clampin et al.(2003)]{2003AJ....126..385C} Clampin, M., et al.\ 
2003, \aj, 126, 385 

\bibitem[Dohnanyi(1969)]{1969JGR....74.2531D} Dohnanyi, J.~W.\ 1969, \jgr, 
74, 2531 

\bibitem[Engelbracht et al.(2007)]{2007PASP..119..994E} Engelbracht, C.~W., 
et al.\ 2007, \pasp, 119, 994 

\bibitem[Giardino et al.(2008)]{2008A&A...490..113G} Giardino, G., Pillitteri, 
I., Favata, F., \& Micela, G.\ 2008, \aap, 490, 113 

\bibitem[Gordon et al.(2005)]{2005PASP..117..503G} Gordon, K.~D., et al.\ 
2005, \pasp, 117, 503 

\bibitem[Habing et al.(1996)]{1996A&A...315L.233H} 
Habing, H.~J., et al.\ 1996, \aap, 315, L233 

\bibitem[Hage \& Greenberg(1990)]{1990ApJ...361..251H} Hage, J.~I., 
\& Greenberg, J.~M.\ 1990, \apj, 361, 251 

\bibitem[Heap et al.(2000)]{2000ApJ...539..435H}
Heap, S.~R., et al.\ 2000, \apj, 539, 435

\bibitem[Henry et al.(1996)]{1996AJ....111..439H} Henry, T.~J., Soderblom, 
D.~R., Donahue, R.~A., \& Baliunas, S.~L.\ 1996, \aj, 111, 439 

\bibitem[Hines et al.(2006)]{2006ApJ...638.1070H} Hines, D.~C., et al.\ 
2006, \apj, 638, 1070 

\bibitem[Jasinta et al.(1995)]{1995A&AS..114..487J} Jasinta, D.~M.~D., 
Raharto, M., \& Soegiartini, E.\ 1995, \aaps, 114, 487 

\bibitem[Jourdain de Muizon et al.(1999)]{1999A&A...350..875J} 
Jourdain de Muizon, M., et al.\ 1999, \aap, 350, 875 

\bibitem[Kalas et al.(2005)]{2005Natur.435.1067K} Kalas, P., Graham, J.~R., 
\& Clampin, M.\ 2005, \nat, 435, 1067 

\bibitem[Kalas et al.(2008)]{2008Sci...322.1345K} Kalas, P., et al.\ 2008, 
Science, 322, 1345 

\bibitem[Krist(2004)]{2004SPIE.5487.1284K} Krist, J.~E.\ 2004, \procspie, 
5487, 1284 

\bibitem[Krist et al.(2005)]{2005AJ....129.1008K} Krist, J.~E., et al.\ 
2005, \aj, 129, 1008 

\bibitem[Krist et al.(2006)]{STinyTIM} Krist, J.~E. 2006, Tiny Tim for 
Spitzer Version 2.0, http://ssc.spitzer.caltech.edu/archanaly/contributed/stinytim/index.html

\bibitem[Lachaume et al.(1999)]{1999A&A...348..897L} Lachaume, R., Dominik, C., Lanz, T., \& Habing, H.~J.\ 1999, \aap, 348, 897

\bibitem[Laor \& Draine(1993)]{laor93} Laor, A.~\& Draine, B.~T.\ 1993, \apj, 402, 441

\bibitem[Lu et al.(2008)]{2008PASP..120..328L} Lu, N., et al.\ 2008, \pasp, 
120, 328 

\bibitem[Mamajek \& Hillenbrand(2008)]{2008ApJ...687.1264M} Mamajek, E.~E., 
\& Hillenbrand, L.~A.\ 2008, \apj, 687, 1264 

\bibitem[Mayor et al.(2004)]{2004A&A...415..391M} Mayor, M., Udry, S., Naef, 
D., Pepe, F., Queloz, D., Santos, N.~C., \& Burnet, M.\ 2004, \aap, 415, 391 

\bibitem[Mo{\'o}r et al.(2006)]{2006ApJ...644..525M} Mo{\'o}r, A., 
{\'A}brah{\'a}m, P., Derekas, A., Kiss, C., Kiss, L.~L., Apai, D., Grady, 
C., \& Henning, T.\ 2006, \apj, 644, 525 

\bibitem[Morales et al.(2009)]{2009ApJ...699.1067M} Morales, F.~Y. et al.\, 2009, \apj, 699, 1067

\bibitem[Nilsson et al.(2010)]{2010arXiv1005.3215N} Nilsson, R., et al.\ 
2010, arXiv:1005.3215 

\bibitem[Perryman et al.(1997)]{1997A&A...323L..49P} Perryman, M.~A.~C., 
et al.\ 1997, \aap, 323, L49 

\bibitem[Plavchan et al.(2009)]{2009ApJ...698.1068P} Plavchan, P., Werner, 
M.~W., Chen, C.~H., Stapelfeldt, K.~R., Su, K.~Y.~L., Stauffer, J.~R., 
\& Song, I.\ 2009, \apj, 698, 1068 

\bibitem[Quillen(2006)]{2006MNRAS.372L..14Q} Quillen, A.~C.\
2006, \mnras, 372, L14

\bibitem[Rebull et al.(2008)]{2008ApJ...681.1484R} Rebull, L.~M., 
Stapelfeldt, K.~R. et al.\ 
2008, \apj, 681, 1484 


\bibitem[Royer et al.(2002)]{2002AA...393.897} Royer, G., Grenier, S., Baylac, 
M.-O., Gomez, A. E., \& Zorec, J. 2002, \aap, 393, 897.

\bibitem[Saija et al.(2003)]{2003MNRAS.341.1239S} Saija, R., Iat{\`i}, 
M.~A., Giusto, A., Borghese, F., Denti, P., Aiello, S., 
\& Cecchi-Pestellini, C.\ 2003, \mnras, 341, 1239 

\bibitem[Sch{\"u}tz et al.(2004)]{2004A&A...424..613S} Sch{\"u}tz, O., 
B{\"o}hnhardt, H., Pantin, E., Sterzik, M., Els, S., Hahn, J., \& Henning, T.\ 
2004, \aap, 424, 613 

\bibitem[Sheret et al.(2004)]{2004MNRAS.348.1282S} Sheret, I., Dent, 
W.~R.~F., \& Wyatt, M.~C.\ 2004, \mnras, 348, 1282 

\bibitem[Smith et al.(2007)]{2007PASP..119.1133S} Smith, J.~D.~T., et al.\ 
2007, \pasp, 119, 1133  

\bibitem[Soderblom(1985)]{1985PASP...97...54S} Soderblom, D.~R.\ 1985, 
\pasp, 97, 54 

\bibitem[Song et al.(2003)]{2003ApJ...599..342S} Song, I., Zuckerman, B., 
\& Bessell, M.~S.\ 2003, \apj, 599, 342 

\bibitem[Song et al.(2004)]{2004ApJ...614L.125S} Song, I., Zuckerman, B., 
\& Bessell, M.~S.\ 2004, \apjl, 614, L125 

\bibitem[Su et al.(2006)]{2006ApJ...653..675S} Su, K.~Y.~L., et al.\ 2006, 
\apj, 653, 675 

\bibitem[Tanner et al.(2009)]{2009ApJ...704..109T} Tanner, A., Beichman, 
C., Bryden, G., Lisse, C., \& Lawler, S.\ 2009, \apj, 704, 109 

\bibitem[Trilling et al.(2008)]{2008ApJ...674.1086T} Trilling, D.~E., et 
al.\ 2008, \apj, 674, 1086 

\bibitem[van Leeuwen(2007)]{2007A&A...474..653V} van Leeuwen, F.\ 2007, 
\aap, 474, 653 

\bibitem[Walker \& Wolstencroft(1988)]{1988PASP..100.1509W} 
Walker, H.~J., \& Wolstencroft, R.~D.\ 1988, \pasp, 100, 1509 

\bibitem[Zuckerman \& Webb(2000)]{2000ApJ...535..959Z} Zuckerman, B., 
\& Webb, R.~A.\ 2000, \apj, 535, 959 

\bibitem[Zuckerman et al.(2001)]{2001ApJ...559..388Z} Zuckerman, B., Song, 
I., \& Webb, R.~A.\ 2001, \apj, 559, 388 

\bibitem[Zuckerman \& Song(2004)]{2004ApJ...603..738Z} Zuckerman, B., 
\& Song, I.\ 2004, \apj, 603, 738 

\end{thebibliography}
\end{document}